\documentclass[twocolumn, secnumarabic, amssymb, nobibnotes, aps, prmaterials, dbfloatfix, superscriptaddress]{revtex4-2}
\usepackage[export]{adjustbox} 
\usepackage{ragged2e}
\usepackage[justification=justified,singlelinecheck=false]{caption}
\DeclareCaptionJustification{justified}{\justifying}
\usepackage{makecell}

\usepackage{chemformula}
\usepackage{siunitx}
\usepackage{amstext}
\usepackage{amssymb}
\usepackage{amsmath}
\usepackage{float}
\usepackage[version=4]{mhchem}
\usepackage{siunitx}
\DeclareSIUnit{\wtpercent}{wt.\%}
\DeclareSIUnit{\atpercent}{at.\%}
\DeclareSIUnit{\sccm}{sccm}
\usepackage{leftindex}
\usepackage{mathtools}

\usepackage{nicefrac,xfrac}
\usepackage{booktabs}
\usepackage{multirow}

\usepackage{color,soul}

\usepackage[labelfont=bf]{caption}
\usepackage{graphicx}
\usepackage{epstopdf}
\DeclareGraphicsExtensions{.eps,.pdf,.png}

\usepackage{subcaption}
\usepackage[colorlinks, allcolors = blue]{hyperref}

\usepackage{leftindex}
\expandafter\let\csname equation*\endcsname\relax
\expandafter\let\csname endequation*\endcsname\relax

\usepackage{amstext}
\usepackage{amssymb}
\usepackage{amsmath}

\usepackage[version=4]{mhchem}

\begin{document}

\title[]{Effective masses, Burstein–Moss shift, and bandgap renormalization in degenerate Al-doped ZnO from broadband ellipsometry and Hall measurements}

\author{S. Mishra} \email[Contact author: ]{smishra@pucp.edu.pe}
\affiliation{Departamento de Ciencias, Secci\'on F\'isica, Pontificia Universidad Cat\'olica del Per\'u, Av. Universitaria 1801, 15088 Lima, Per\'u}

\author{L. A. Enrique}
\affiliation{Departamento de Ciencias, Secci\'on F\'isica, Pontificia Universidad Cat\'olica del Per\'u, Av. Universitaria 1801, 15088 Lima, Per\'u}
\affiliation{Department Perovskite Tandem Solar Cells, Helmholtz-Zentrum Berlin für Materialien und Energie GmbH, 12489 Berlin, Germany}

\author{D. Cespedes}

\author{E. Perez}
\affiliation{Departamento de Ciencias, Secci\'on F\'isica, Pontificia Universidad Cat\'olica del Per\'u, Av. Universitaria 1801, 15088 Lima, Per\'u}

\author{E. Serquen}
\affiliation{Departamento de Ciencias, Secci\'on F\'isica, Pontificia Universidad Cat\'olica del Per\'u, Av. Universitaria 1801, 15088 Lima, Per\'u}

\author{F. Bravo}
\affiliation{Departamento de Ciencias, Secci\'on F\'isica, Pontificia Universidad Cat\'olica del Per\'u, Av. Universitaria 1801, 15088 Lima, Per\'u}

\author{P. Llontop}
\affiliation{Departamento de Ciencias, Secci\'on F\'isica, Pontificia Universidad Cat\'olica del Per\'u, Av. Universitaria 1801, 15088 Lima, Per\'u}
\affiliation{Materials to Optoelectronic Devices Group and Institute of Complex Molecular Systems, Eindhoven University of Technology, Eindhoven 5600 MB, The Netherlands}

\author{F. Ruske}
\affiliation{Department Perovskite Tandem Solar Cells, Helmholtz-Zentrum Berlin für Materialien und Energie GmbH, 12489 Berlin, Germany}
\author{L. Korte}
\affiliation{Department Perovskite Tandem Solar Cells, Helmholtz-Zentrum Berlin für Materialien und Energie GmbH, 12489 Berlin, Germany}
\author{J. A. Guerra}\email[Contact author: ]{guerra.jorgea@pucp.edu.pe}
\affiliation{Departamento de Ciencias, Secci\'on F\'isica, Pontificia Universidad Cat\'olica del Per\'u, Av. Universitaria 1801, 15088 Lima, Per\'u}

\date{\today}

\begin{abstract}
A systematic methodology is established to determine the electron and hole effective masses in degenerate transparent conducting oxides. Demonstrated on Al-doped ZnO thin films, the approach performs a simultaneous global fit of the optical bandgap and squared plasma energy as functions of carrier concentration, explicitly accounting for band nonparabolicity and careful modeling of the material’s optoelectronic properties. Accurate optical bandgap and plasma frequency are determined by an  Elliott-Band Fluctuations fundamental absorption model and a modified Sernelius optical description of the free carrier response, respectively. By analyzing a range of samples with carrier concentrations covering the Mott transition, we reveal pronounced conduction band nonparabolicity features. We show that neglecting valence band contributions produces systematic errors in both the extracted effective mass and bandgap shift, whereas the parabolic band approximation underestimates the carrier-induced bandgap shift by up to $\sim0.35$ eV.  The electron effective mass is found to vary in the range of $0.095$ to $0.42\,m_e$ over the studied carrier-concentration range, with a nonparabolicity parameter $C^{-1}$ of $3.85$ eV, within the random phase approximation. The average hole effective mass is determined to be at $0.32\,m_e$.  This methodology provides a unified route to extracting bandstructure parameters, disentangling band-filling and many-body effects and is readily applicable to other degenerate transparent semiconductors.

\end{abstract}
\maketitle
\section{Introduction}
 Renewed interest in  electronic transport and thermoelectric performance in degenerate semiconductors has highlighted the critical role of electronic band structure and nonparabolicity \cite{Mecholsky,Hosono2017,hartnagel1995,Erkan,Ruske2012ZnO,KEllmer_2001,ABDOLAHZADEHZIABARI2012,Walsh,Feneberg16,Feneberg14,Whalley}. Beyond transport phenomena, the bandgap remains a fundamental parameter for understanding optical properties in transparent conducting oxides (TCOs), particularly in governing parasitic absorption and infrared transparency \cite{Schultes,Berggren}. Among TCOs, Al-doped ZnO (AZO) has emerged as an attractive alternative to indium-based systems owing to its earth abundance, non-toxicity, and stability under hydrogen-plasma conditions \cite{Hosono2017,hartnagel1995,Erkan,Ruske2012ZnO,KEllmer_2001,Kim_2008}. However, optimizing TCOs requires managing a fundamental transport trade-off governed by the electron effective mass ($m_c^*$). Elevating carrier concentration ($n_e$) enhances electrical conductivity but simultaneously raises the plasma frequency ($\omega_p \propto \sqrt{n_e/m_c^*}$), increasing free-carrier absorption in the near-infrared and reducing device efficiency in applications such as tandem solar cells \cite{Schultes}. Despite their technological relevance, fully elucidating the microscopic mechanisms limiting charge transport and carrier mobility (\( \mu_e \)) remains an ongoing challenge, underscoring the need for continued fundamental studies \cite{Singh2024,Young,nano14070591,Enrique_2025}.

The effective mass approximation provides a useful framework for describing diverse material properties, including optical properties such as excitonic binding energies, transport properties such as polaron radii and carrier mobilities, and shallow defect states \cite{Whalley,Kevin_2025PRM}. These quantities can be compared with first‑principles band‑structure calculations, an especially important benchmark for wide bandgap semiconductors, where absorption edges at photon energies  above,  $\sim 3 \,\text{eV}$ limit access to higher‑order band-to-band transitions by conventional spectroscopy \cite{Shirayama}. Despite its importance, however, the experimental determination of electron and hole effective masses remains challenging. A wide range of experimental and theoretical approaches have been employed to determine the effective masses and nonparabolicity parameters of ZnO and, in a lesser extent of AZO. Experimental determination of the effective mass depend strongly on carrier concentration, sensitivity to directional band dispersion, and instrumental uncertainties, while theoretical predictions rely on the choice of exchange–correlation functions, strain conditions, and many‑body corrections. Techniques such as Cyclotron resonance and de Haas–van Alphen demand high mobilities ($\sim 300\,\mathrm{cm^2\,V^{-1}\,s^{-1}}$), strong magnetic fields \((B)\), and long relaxation times \((\sim 10^{-15}\,\mathrm{s})\) to satisfy \(\mu_e B \gg 1\), conditions rarely met in polycrystalline TCOs \cite{Young,IMANAKA2001,Button,Oshikiri,OSHIKIRI2001}. Four‑coefficient methods provide out‑of‑plane (perpendicular) effective masses and are highly sensitive to multi-temperature parameter uncertainties  \cite{Young,Kim_2008}. Optoelectronic analyses combining for instance, spectroscopic ellipsometry with Hall effect measurements yields a mixed response of the parallel and perpendicular effective mass components \cite{Kim_2008,GoriPRB10}. Photoelectron spectroscopy  requires careful treatment of the bandgap and remains challenging due to the difficulty in disentangling Fermi-level variations from doping-induced bandgap renormalization, in addition to other experimental limitations \cite{PhysRevLett.100.167402,Jia,Lim2012,Preston2008}. Furthermore, directional band parameters determined from single-crystal anisotropic probes (such as photoreflectance) do not straightforwardly translate to heavily impurity-doped polycrystalline thin films, where crystallite orientations, defect scattering, and grain boundaries average out single-crystal symmetries \cite{IMANAKA2001,Button,Oshikiri,OSHIKIRI2001,SYRBU2004}. From a theoretical perspective, both \(k \cdot p\) perturbation theory and density functional theory (DFT) calculations remain highly sensitive to the choice of exchange-correlation functional, where semi-local approximations systematically underestimate the bandgap as well as assumptions regarding strain and impurity-induced lattice distortions \cite{Lambrecht,Xu,OSHIKIRI2001,Klingshirn,klingshirn2010,Sikam_2020,Tripathi2020,ShadangiIOP2016,CARDONA1963}.

Previous studies, mainly on ZnO, show poor agreement of electronic band parameters, leaving open questions regarding their true fundamental values. Here, we attempt to shed some light by employing an optoelectronic approach that integrates optical spectroscopy with room-temperature transport measurements and careful modeling of the fundamental absorption and free-carrier optical response. While similar strategies have been applied to Al  and Ga doped ZnO films and other TCOs \cite{Fujiwara,RUSKE2009,Feneberg16}, systematic investigations of conduction band nonparabolicity remain scarce and the reported values vary substantially between studies \cite{RUSKE2009,Fujiwara,Kim_2008,Bikowski}.  Often, the nonparabolicity parameter is extracted by fixing the band edge electron mass, preventing independent determination of both quantities \cite{RUSKE2009,Bikowski}. Alternatively, models rely either on simplified parabolic approximations or on nonparabolic approaches that assume a flat valence band, neglecting hole contributions to both the Burstein–Moss shift (BMS) and bandgap renormalization (BGR) alongside oversimplified screening models for BGR \cite{GUPTA198933,Hamberg,Berggren,Sernelius,Feneberg16,Walsh,ABDOLAHZADEHZIABARI2012,Asghar,RUSKE2009,Fujiwara,Bikowski}. The  optical bandgap  determination introduces a further source of systematic uncertainty when disorder-induced tail states and excitonic contributions near the absorption edge are neglected \cite{Asghar,Enrique_2025,Makino_2005,Makino2002}. This is particularly relevant for ZnO-based systems, where excitonic absorption features have been reported to persist at carrier concentrations approaching $10^{20}\,\mathrm{cm^{-3}}$, consistent with Mahan-type excitonic effects under strong electrostatic screening \cite{Enrique_2025,Makino_2005, Mahan, Yamamoto,  Papadimitriou, Kabir_2019, Asghar, ABDOLAHZADEHZIABARI2012, SPADONI2015514}.

In this work, we address these limitations by investigating sputtered AZO thin films across carrier concentrations spanning the Mott insulator-to-metal critical threshold,  $n_c \sim 4.85\times10^{19}\,\mathrm{cm^{-3}}$, enabling the analysis of the optical response through the transition to a degenerate system. The complex dielectric function \((\tilde{\varepsilon}=\varepsilon_1+i\varepsilon_2)\), is determined by combining broadband (\qty{200} to \qty{3500}{\nm}) variable angle spectroscopic ellipsometry and spectrophotometry in a self-consistent point-by-point procedure without imposing predefined optical-dispersion models \cite{Guerra17,Tejada,Enrique_2025}. The fundamental absorption edge is analyzed using the Elliott-Band Fluctuations (EBF) model, explicitly accounting for excitonic  and disorder-induced Urbach-tail contributions, yielding the optical bandgap \( (E_g^{\text{opt}}) \), Urbach energy \( (E_U) \), and  excitonic binding energy \( (E_b) \) \cite{Makino_2005,kevin25, Enrique_2025,Zhang23MDPI}. Concurrently, the free carrier response is described with a model including impurity scattering and the polar lattice response of ZnO \cite{PFLUG, Enrique_2025,RUSKE2009}, from which the plasma ($\omega_p$) and damping ($\omega_\tau$) frequencies are retrieved. The central element of our analysis is a global fit of the carrier-concentration dependence of both the optical bandgap  \( (E_g^{\text{opt}}) \) and square plasma energy $(\hbar\omega_p)^2$, enabling  simultaneous determination of the band edge electron effective mass (${m_{c_0}^\ast}$), mean hole effective mass (${m_{v_0}^\ast}$), nonparabolicity parameter ($C$) and intrinsic bandgap (${E_{g_0}}$), while explicitly accounting for thermal and disorder induced states in degenerately doped semiconductors.
The BMS is described using the Kane quasilinear band dispersion \cite{KANE1957249}, allowing ${m_{c_0}^\ast}$ and $C$ to be determined without fixing either parameter a priori. The BGR shift is calculated from carrier-induced self-energy corrections in both conduction- and valence- bands, thereby retaining the contribution of the valence band to the bandgap shift. Carrier screening is evaluated using both the plasmon pole (PPA) and the random phase (RPA) approximations. While the PPA models the dielectric screening via a single effective plasmon mode, the RPA retains the full dynamical carrier response and provides a more comprehensive description of screening in the degenerate regime. This framework therefore separates the contributions of band filling, many-body bandgap renormalization, band nonparabolicity, and absorption edge broadening to the carrier concentration dependent optical response, providing a stringent determination of the fundamental band parameters of polycrystalline AZO thin films. The analysis can be readily extended to other transparent conducting oxides and degenerate transparent semiconductors.

\section{Methodology}
\begin{figure*}[t]
\centering    
\includegraphics[width=\textwidth]{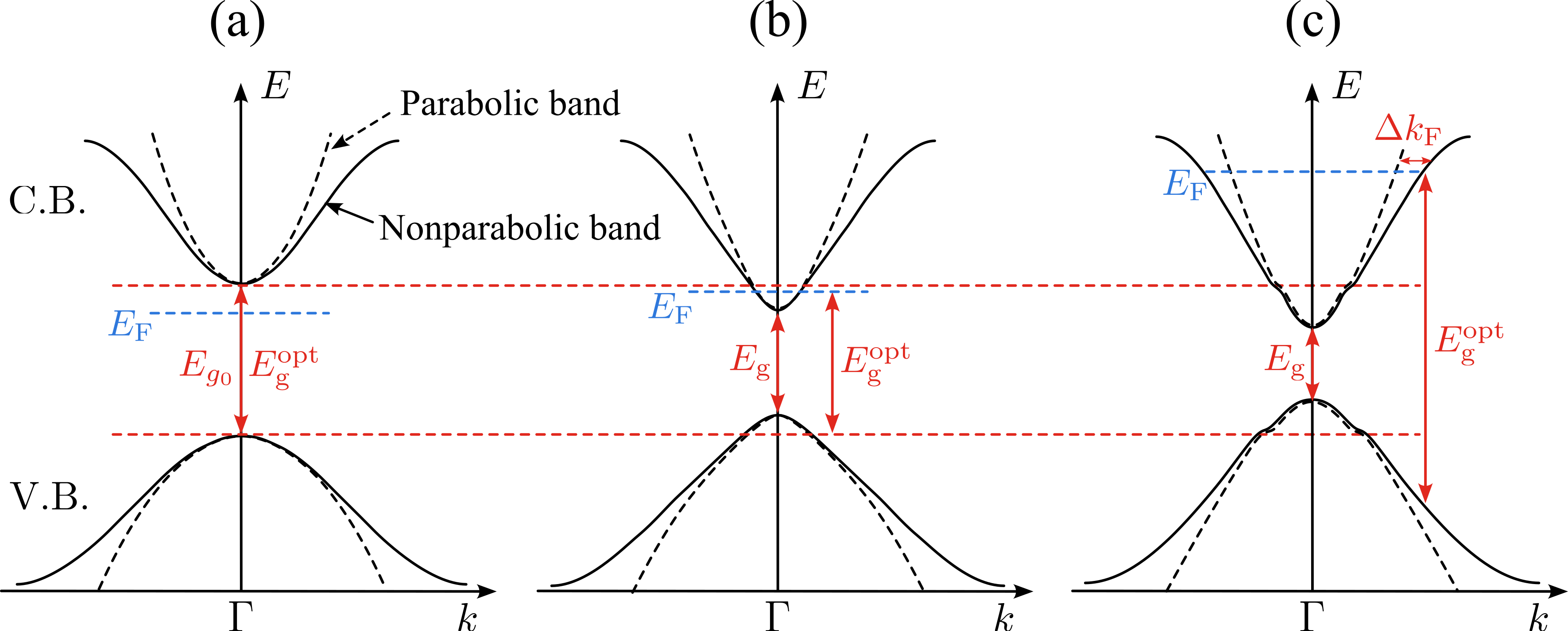}
\caption{
Schematic representation of the Fermi level $(E_F)$ position depicting  n-type character for different carrier concentrations. Nondegenerate ZnO showing the intrinsic bandgap \( E_{g_0} \) with the Fermi level located close to the conduction band (C.B.) edge due to intrinsic and/or extrinsic  donor defects (a). Moderate carrier concentration just above the Mott critical concentration, \( n_c \), elevating the Fermi level, and resulting in a subtle Burstein-Moss shift (BMS) and  slight perturbation arising from bandgap renormalization (b). At high carrier concentrations, \( E_F \) enters well into the degenerate regime beyond the $n_c$ threshold, highlighting the Fermi wave vector deviation  $\Delta k_F$  when nonparabolic features are neglected (c). Additionally, enhanced perturbations arising from many-body interactions are shown to shrink the  fundamental bandgap \( (E_g) \) of  AZO, while simultaneously leading to an enhancement of the optical bandgap $(E_g^{\text{opt}})$.
}

\label{Figure1}
\end{figure*}

In nondegenerate semiconductors, the optical bandgap \( E_g^{\text{opt}} \), corresponding to the fundamental bandgap, coincides with the energy difference between the conduction band minimum and the valence band maximum i.e., the intrinsic bandgap ($E_{g0} $). However, under conditions of carrier accumulation or doping-induced degeneracy, the lowest conduction band states near the band edge become partially occupied. For ZnO, whose conduction band minimum is located at the $\Gamma$ point, this occupation suppresses electronic transitions into these states. Consequently, the lowest-energy allowed direct transitions occur between occupied states in the valence band and unoccupied states at a finite $k$ electronic wave vector, away from the $\Gamma$ point. The carrier occupation near the Fermi level $E_F$, is governed by the Fermi–Dirac distribution, with thermal broadening on an energy scale in the order of $k_{\beta}T$, with \( k_\beta \) and \( T \) the Boltzmann constant and the absolute temperature, respectively. Optically induced electronic transitions contributing to the absorption are not restricted to the latter narrow energy range and extend over a broader range of energies and  $k$ states. This suppression of electronic transitions  near the band edges increases the energy threshold for an interband electronic transition to take place.
Thus, \( E_g^{\text{opt}} \) increases departing from $E_{g_0} $. The evolution of this phenomenon is illustrated in Fig. \ref{Figure1}. This occupation-induced blue shift of the absorption edge onset is referred as the Burstein-Moss shift (BMS) \cite{Schleife,glutsch2004}, which accounts for the carrier concentration–dependent increase of \( E_g^{\text{opt}} \). Concurrently, many-body interactions, including exchange and Coulomb interactions among free carriers as well as carrier–impurity interactions, lead to bandgap narrowing. These effects induce downward and upward shifts of the conduction band minimum and valence band maximum, respectively, thereby reducing the fundamental bandgap $(E_g)$ of the material. This phenomenon, known as bandgap renormalization (BGR), is illustrated in  Fig. \ref{Figure1}(b) and \ref{Figure1}(c) for moderate and high carrier concentrations, respectively  \cite{Sernelius}.

The net  bandgap  shift reflects the interplay between BMS and BGR, and has been extensively modeled using quasiparticle dispersion relations, often under the parabolic band approximation in the moderate carrier concentration regime \cite{GUPTA198933,Berggren,Hamberg,Sernelius}. For the sake of consistency and comparison purposes, in the following subsections we address the BMS and BGR in the parabolic band approximation. Subsequently, this is addressed for nonparabolic band approximations, in the framework of the models developed by Pisarkiewicz \cite{PISARKIEWICZ1989217,Pisarkiewicz} and Nilsson \cite{Nilsson,BLAKEMORE19821067, KEllmer}.

\subsection{Burstein–Moss Shift and Bandgap Renormalization in Parabolic Bands}

The optical bandgap can  be calculated by the energy difference between the available occupied and unoccupied states, i.e., at the Fermi wave vector ($k_F$), in the conduction \( (E_c) \) and valence \( (E_v) \) bands by
 
\begin{equation}
    E_g^{\text{opt}}=E_c\left(k_{F,}\omega\right)-E_v(k_{F,}\omega).
\end{equation}

To account for electron–electron and electron–impurity scattering, the unperturbed band dispersions are replaced by quasiparticle dispersions \cite{Hamberg}. Under these conditions, the valence and conduction energy bands take the form 

\begin{subequations} \label{eq4}
\begin{equation}
    E_v\left(k_{F,}\omega\right)=\frac{-\hbar^2k^2_F}{2m_{v_0}^*}+\hbar\sum\nolimits_v(k_F,\omega),
    \label{eq4(a)}
\end{equation}

and,
\begin{equation}
E_c\left(k_{F,}\omega\right)=E_{g0}+ \frac{\hbar^2k^2_F}{2m_{c_0}^*} + \hbar\sum\nolimits_c(k_F,\omega) .
  \label{eq4(b)}
\end{equation}
\end{subequations}

Here, $\hbar \sum_v$  and $\hbar\sum_c$  are the self-energy interaction terms due to electron-electron and electron-impurity scattering in the valence and conduction bands, respectively. $m_{c_0}^\ast$ and  $m_{v_0}^\ast$ are the near band edge electron and hole effective masses, respectively.

Then, \( E_g^{\text{opt}} \) can be written in terms of the BMS and BGR components, 
\begin{equation}
E_g^{\text{opt}} = E_{g_0} + \Delta E_{g}^{\text{BMS}}- \Delta E_g^{\text{BGR}},
\label{eq3}
\end{equation}
with the BMS component within the parabolic band approximation given by

\begin{equation}
    \Delta E_{g}^{\text{BMS}}=\dfrac{\hbar^2k_F^2}{2m^*_{c_0v_0}}=\dfrac{\hbar^2(3\pi^2n_e)^{2/3}}{2m^*_{c_0v_0}}.
    \label{eq4}
\end{equation}

Here, the dependency of the Fermi wave vector ($k_F$) on the  carrier concentration ($n_e$), \mbox{$k_F= \left(3\pi^2n_e\right)^{1/3}$}, is used.  $m_{c_0v_0}^\ast$ denotes the reduced effective mass in  terms of the near band edge electron and hole effective masses. 
\begin{equation}
    \dfrac{1}{m_{c_0v_0}^*}=\dfrac{1}{m_{c_0}^*}+\dfrac{1}{m_{v_0}^*}.
\end{equation}

On the other hand, the BGR component is given by 
\begin{equation}
\Delta E_g^{\text{BGR}} = \hbar \sum\nolimits_v(k_F, \omega) - \hbar \sum\nolimits_c(k_F, \omega).
\notag
\end{equation}
The sign convention accompanying  $\Delta E_g^{BGR}$ in Eq. (\ref{eq3}) is arbitrary, so the final choice is purely a matter of convention. Here, we choose a negative sign as a reminder that BGR shrinks the bandgap. 

The magnitude of the self-energies is calculated in accordance with the method described by Berggren \cite{Berggren}, either within the PPA or RPA approximations. The self-energy terms were originally derived and tested for silicon (\ch{Si}) and germanium (\ch{Ge}). However, these expressions have since been successfully applied to other materials, including oxide semiconductors, as reported in subsequent studies \cite{GUPTA198933,Hamberg,Sernelius,Feneberg16,Feneberg14}.

Electron-electron and electron-impurity scattering are additive processes, thus, their contribution to the BGR can be expressed as

\begin{subequations} \label{eq6}
\begin{equation}
\hbar \sum\nolimits_{c}(k_F, \omega) = \hbar \sum\nolimits_{c}^{ee}(k_F, \omega) + \hbar \sum\nolimits_{c}^{ei}(k_F, \omega),
\end{equation}
for the conduction band, and
\begin{equation}
\hbar\sum\nolimits_{v}(k_F, \omega) = \hbar \sum\nolimits_{v}^{ee}(k_F, \omega) + \hbar \sum\nolimits_{v}^{ei}(k_F, \omega).
\end{equation}
\end{subequations}
for the valence band.

Despite accounting for additive electron-electron and electron-impurity interactions, in the framework of the parabolic band approximation for a highly degenerate semiconductor, the BGR effects can be biased by up to $0.5~\mathrm{eV}$ when neglecting nonparabolic features. This bias is further heightened when the hole effective mass is neglected, as has been done in previous reports \cite{Walsh, Sernelius,GUPTA198933,ABDOLAHZADEHZIABARI2012}. Further details on the latter bias calculations can be found in SM Fig. S8. These effects highlight the need to incorporate wavevector ($k$) dependent corrections.

\subsection{Nonparabolic Band Dispersion Models}
 Near the band edge and around the Mott critical concentration, the deviation between parabolic and nonparabolic energy band dispersions can be assumed  to be minimal. However, once the carrier concentration exceeds the  threshold of the Mott critical concentration, the system enters deep into the degenerate regime, marked by a discernible blue shift in the fundamental optical absorption edge. In this regime, $E_F$ resides well within the conduction band, which brings $k_F$ well apart from the  $\Gamma$ point. As a consequence, electronic transitions occur within the nonparabolic regime of the band structure, as shown in Fig.~\ref{Figure1}(c), causing the parabolic band approximation to systematically misestimate both the Fermi level and the Fermi wave vector shifts. The Kane type quasilinear dispersion relation, derived from the two-band $k\cdot {p}$ formalism can be used to describe the energy band dispersion relation observed in degenerate semiconductors \cite{KANE1957249,Cardona,Shokhovets}. This approach captures the nonparabolic $k$-dependent corrections to the conduction band arising from the strong coupling with the valence band.  In this framework, the energy band dispersion relation, with respect to the conduction band edge, is expressed as:
\begin{equation}
    E(k)=\frac{E_g}{2}+\sqrt{\left(\frac{E_g}{2}\right)^2+4t^2\sin^2\left(\frac{ka}{2}\right)}.
    \label{eq8}
\end{equation}

Here, $a$ is the lattice constant or discretizing length, and \emph{t} corresponds to the off-diagonal coupling matrix element arising from band mixing in the two-band model. This coupling strength can be estimated using the  following relation \cite{DattaQuantumPhenomena,AshcroftMermin1976}: 
\begin{equation}
    t^2=\dfrac{\hbar^2{E_g}}{2a^2m^*_{c_0 v_0}}.
    \label{eq9}
\end{equation}
Substituting \emph{t}  in Eq. (\ref{eq8}) and solving for the near band edge regime $(ka \ll 1)$ with constraints, $E(k) \geq {E_g}
$, i.e., restricted to the conduction band only, gives
\begin{equation}
    E_c(1+C E_c)=\dfrac{\hbar^2k^2}{2m^*_{c_o}}.
    \label{eq10}
\end{equation}
Here, $m_{c_0}^\ast$  is the electron effective mass at the  bottom of the conduction band and  $C$ is a positive  constant known as the band nonparabolicity parameter which approximately satisfies \( C \sim \nicefrac{1}{E_{g0}}\). A detail derivation of Eq. (\ref{eq10}) is provided in Section S1 of the Supplementary Material (SM). Notice that for $C = 0$, Eq. (\ref{eq10}) simplifies to the parabolic band relation.

From the quasi-momentum ($\hbar k$) and the velocity ($v$) of an electron wave packet relation, $mv = \hbar k$, \cite{Whalley} the electron effective mass, often referred to as transport effective mass  can be generally expressed in terms of the energy band dispersion relation by,
 
\begin{equation}
    \dfrac{1}{m^*_c}=\dfrac{1}{\hbar^2k}\dfrac{\partial E_c (k)}{\partial k}.
\end{equation}
Thus, by differentiating Eq. (\ref{eq10}), and evaluating at Fermi surface with $k_F = (3\pi^2n_e)^{1/3} $, a carrier dependent (transport) effective mass is obtained, i.e. 
 \begin{equation}
m_c^*(k_F)=m_{c_0}^*\left(1+2CE_c(k_F)\right).
 \label{eq12}
\end{equation}

In the following subsections, we briefly describe the models of Pisarkiewicz \cite{PISARKIEWICZ1989217,Pisarkiewicz} and Nilsson \cite{Nilsson}, which provide ways to calculate the energy-dependent effective mass in the nonparabolic regime. Within the framework of this study, it is useful to express this relation as a function of the carrier concentration $n_e$. Therefore, the corresponding formulation is recast in terms of $n_e$.

\subsubsection*{Nonparabolic Band Model by Pisarkiewicz}
Pisarkiewicz \emph{et al.} derived a nonparabolic dispersion relation under the assumption of complete carrier degeneracy and spherically symmetric energy bands. The integration of the occupied density of states is then performed by approximating the Fermi distribution to a step function \cite{PISARKIEWICZ1989217,Pisarkiewicz}. Through this formulation, Eq.(\ref{eq12}), becomes

\begin{equation}
    \frac{m^*_c}{m^*_{c_0}} = \sqrt{1+2C\dfrac{\hbar^2}{m_{c_0}^*}(3\pi^2n_e)^{2/3}} .
    \label{eq13}
\end{equation}
The Fermi energy with respect to the conduction band edge can then be retrieved from Eq. (\ref{eq10}) evaluated at the Fermi wavevector $k_F$.
 
\begin{align}
    E_c(k_F) &= \nonumber \\
    E_F &=\frac{1}{2C}\left(\sqrt{1+2C\dfrac{\hbar^2}{m_{c_0}^*}(3\pi^2n_e)^{2/3}} -1\right)
    \label{eq14}
\end{align}

The Pisarkiewicz Model offers a ready-to-use formula. However, it does not take into account thermal  occupation effects nor the impact of impurity and disorder-induced band tailing on the density of states $g(E)$, which are critical when evaluating the conduction band edge in heavily doped semiconductors. These effects were subsequently addressed  by Nilsson \cite{Nilsson}, Blakemore \cite{BLAKEMORE19821067} and, Ellmer \cite{KEllmer}. As Nilsson’s model provides a foundational framework for band nonparabolicity, with further refinements reported in subsequent works, we consistently employ Nilsson’s relation throughout this study.

\subsubsection*{Nonparabolic Band Model by Nilsson}

Nilsson proposed a model for the conduction band in which the integration over the occupied density of states is reformulated in terms of the Fermi–Dirac integral $F_{1/2}(\eta)$, where  $\eta = \frac{E_F - E_c}{k_\beta T}$ denotes the reduced Fermi energy. The integral is expressed as $ F_{1/2}(\eta) = n_e/N_c $, where $ N_c $, is the conduction band effective density of states, thereby serving as a normalized occupation parameter. The reduced Fermi energy $(\eta)$ was empirically constructed by introducing two correction functions, designed to be broadly applicable at room temperature, as well as in a wide range of temperatures and carrier concentrations. This model was further substantiated in ZnO by Ellmer, demonstrating its successful applicability under the aforementioned experimental conditions \cite{BLAKEMORE19821067,Nilsson,ABDOLAHZADEHZIABARI2012,KEllmer}.

Based on the work of Nillson, Ellmer \cite{KEllmer} introduced an analytical expression for the electron effective mass ($m^*_c$) as a function of carrier concentration, incorporating thermal and impurity effects on the band edge, 
\vspace{-1em}
\begin{multline}
\frac{m_c^\ast}{m_{c0}^\ast} = 1 + 2C{k_\beta}T \Bigg(
\frac{\log\left( \sfrac{n_e}{N_c} \right)}{1 - \left( \sfrac{n_e}{N_c} \right)^2} \\
+ \frac{\left( 3\pi^{0.5} \left( \sfrac{n_e}{4N_c} \right)^{2/3} \right)}{
1 +  \left[ 0.24 + 1.08\left( 3\pi^{0.5} \left( \sfrac{n_e}{4N_c} \right)^{2/3} \right) \right]^{-2} }
\Bigg),
\label{eq15}
\end{multline}
with the conduction band effective density of states (\( N_c \)), given by
\begin{equation}
 N_c={2\left(\frac{m_{c_0}^\ast k_\beta T}{{2\pi}\hbar^2}\right)}^{3/2}.
 \label{eq16}  
\end{equation}

Like before, the Fermi energy, with respect to the conduction band edge, can be calculated directly from Eq.  (\ref{eq10}), by evaluating the conduction energy band dispersion at $k_F$, i.e., $E_c(k_F)=E_F$. To assess the influence of the assumed dispersion band model, we employed the Pisarkiewicz and Nilsson nonparabolic models to describe the conduction band  and extract the electron and hole effective masses.

\subsection{Burstein–Moss Shift and Bandgap Renormalization in Nonparabolic Bands}
We now evaluate the BMS, analogous to Eq.~(\ref{eq4}), using  nonparabolic band models described in the previous subsection B. In contrast to several previous treatments, the valence band contribution is explicitly included in the reduced effective mass, $m_{cv}^\ast$. Neglecting the valence band edge contribution leads to a systematic underestimation of both the self-energy corrections and BMS by \qty{0.4}{eV}  as shown in Fig.~S8  \cite{Feneberg16,Feneberg14}. In addition this inclusion further enables the extraction of the hole effective mass.

By adopting a nonparabolic conduction band and a parabolic valence band edge (NP(CB)–P(VB)), the BMS can be expressed as

\begin{equation}
\Delta E_{g}^{\mathrm{BMS}}(n_e) = \frac{1}{2C}\left(\frac{m_c^\ast(n_e)}{m_{c_0}^\ast}-1\right)\\ + \frac{\hbar^2}{2 m_{v_0}^\ast}(3\pi^2n_e)^{2/3},
\label{eq17}
\end{equation}
where the carrier concentration dependent electron effective mass $m_c^\ast(n_e)$ can be described either by Pisarkiewicz's (Eq. \ref{eq13}) or Nilsson's (Eq. \ref{eq15}) relations.

However, both band nonparabolicity can also be considered via a joint nonparabolic band (JNPB) dispersion \cite{KANE1957249} defined as
\begin{equation}
    E_{cv}(1+CE_{cv})=\frac{\hbar^2k_F^2}{2m_{c_0v_0}^*} .
    \label{eq18}
\end{equation}
Here, the nonparabolicity curvature parameter $C$ can be related to the degree of admixture of $s$-like CB states and $p$-like VB states  with \( C \sim \nicefrac{1}{E_{g0}}\). 

Calculations employing both NP(CB)-P(VB) and JNPB approximation yield nearly identical effective mass values and bandgap shifts, indicating weak sensitivity to valence band nonparabolicity in the degenerate n-type regime. Therefore, the subsequent analysis follows the NP(CB)-P(VB) approximation for simplicity and to facilitate comparison of the resulting averaged valence band effective mass and nonparabolicity parameter with previous reports. Nevertheless, results using a JNPB can be found in SM Fig. S5, for comparison purposes.

Concerning the BGR, an accurate evaluation of the electron–electron ($e–e$) and electron–ionized impurity ($e–i$) self-energy contributions requires additional material-dependent parameters~\cite{Feneberg16}. Accordingly, the conventional expressions for the effective Bohr radius, Fermi wave vector, and Thomas–Fermi screening length are modified by introducing the carrier concentration-dependent effective mass, $ m^*_c(n_e)$, thereby accounting for band nonparabolicity. The resulting expressions are given in Eqs.~\eqref{eq20}–\eqref{eq22}.

The Bohr radius ,
\begin{equation}
    a^*(n_e)=\frac{4\pi\varepsilon_0\varepsilon_S\hbar^2}{e^2 m_c^*(n_e)}.
\label{eq20}
\end{equation}
The Fermi vector, 
\begin{equation}
    k_F(n_e )=\sqrt{\frac{2 m_{cv}^*(n_e)\Delta E_{g}^{BMS}(n_e)}{\hbar^2}}.
\label{eq21}
\end{equation}

The Thomas-Fermi screening length, in its approximate dependence on the Fermi vector ~\cite{Feneberg16,Berggren,Sernelius,Hamberg},
\begin{equation}
   k_{TF}(n_e)=\sqrt{\frac{4k_F(n_e)}{\pi a^*(n_e)}}.
\label{eq22}
\end{equation}
Here, $\varepsilon_0$ is the vacuum permittivity and  $\varepsilon_S$ is the static dielectric constant of ZnO, with the latter taking a value of 8.65 \cite{Jianguo,Bikowski,Ellmer2008,Sernelius}.

The $e–e$ and $e–i$ self-energy contributions are calculated using the conventional expressions of Berggren \cite{Berggren}, with the associated parameters reformulated to incorporate the effects of band nonparabolicity. The modified equations are provided in section S1 of the SM. It should be stressed that all these calculations pertain to homogeneous systems. Here, the valence band contribution is considered, and for simplicity the band structure factor ($\gamma$),  is assumed to be unity ($\gamma =1$) \cite{Jianguo}. In addition, anisotropy is neglected in the description of dielectric screening, as its influence is generally weak and has been shown to be negligible even in strongly anisotropic systems such as in  n-type many-valley semiconductors (Si, Ge), as demonstrated by Berggren \textit{et al.} \cite{Berggren}. Consistently, optical measurements on single-crystal ZnO report only minor anisotropy effects up to \qty{12}{eV} in the energy range from \SIrange{2.5}{32}{\electronvolt} \cite{GoriPRB10}. Although more pronounced anisotropic features may emerge at higher energies, these lie well above the  relevant spectral region to the present analysis. Therefore, anisotropy is not expected to influence the extraction of the optical parameters of interest in this study.
\vspace{-1em}
\subsection{Global Fitting Methodology}
To achieve a physically consistent determination of the effective mass parameters, a global fitting approach was adopted. The procedure simultaneously accounts for the carrier-concentration dependence of the optical bandgap $E_g^{\mathrm{opt}}$, and the squared plasma energy $(\hbar\omega_p)^2$, given by  Eq. (\ref{eq3}) and  Eq. (\ref{eq26}) respectively. Both quantities depend on the nonparabolic electron effective mass ($m_{c}^\ast$), while $E_g^{\mathrm{opt}}$ additionally contains the valence band contribution. Therefore, a global fit provides  complementary constraints on the band parameters that are not accessible  from the $(\hbar\omega_p)^2$ dependence alone. The squared scale of the plasma energy is chosen since it is on the same order of magnitude as the optical bandgap.
The global fit is performed by minimizing the following reduced unbiased $\chi^2$ global estimator:

\begin{widetext}
\begin{equation}
\chi^2(m_{c_0}^\ast,C,m_{v_0}^\ast,E_{g_0}) =
\frac{\displaystyle \sum_{i=1}^{N} \left( E^\text{opt}_{g\,i} - E^\text{opt}_{g}(n_i; m_{c_0}^\ast, C, m_{v_0}^\ast, E_{g_0}) \right)^2}{\sqrt{N-4-1}}
+ \frac{\displaystyle \sum_{j=1}^{M} \left( \left( \hbar \omega_p \right)^2_j - \left( \hbar \omega_p \right)^2(n_j; m_{c_0}^\ast, C) \right)^2}{\sqrt{M-2-1}}
,
\label{eq23}
\end{equation}
\end{widetext}

Here, N and M are the number of experimental data points for the optical bandgap $( E^\text{opt}_{g\text{ }i})$ and squared plasma energies $(\hbar\omega_p)^2_j$, respectively. The fitting parameters are the band edge electron effective mass $m_{c0}^{*}$, nonparabolicity parameter $C$, Intrinsic bandgap  $E_{g0}$, and mean hole effective mass $m_{v0}^{*}$.
The best-fitted parameters are identified from the minimum of $\chi^2$ over the allowed parameter space. Parameter uncertainties are subsequently determined from confidence limits of the global fit \cite{Guerra17,Press1992}.

To assess the sensitivity of the extracted parameters to the assumed energy band dispersion, the global procedure is applied independently using the Pisarkiewicz and Nilsson relations defined in Eqs.~(\ref{eq13}) and (\ref{eq15}). The best-fitted results are presented in Fig.~\ref{Figure3}, whereas the corresponding independent direct fits are provided in SM Fig.~S4.

 \begin{table*}[t]
\caption{\label{table1}
Summary of  carrier concentrations ($n_e$) and electron mobilities ($\mu_{\text{e}}$) obtained from Hall effect measurements. Plasma frequency ($\omega_p$) and damping frequency ($\omega_{\tau}$) were retrieved after fitting  NIR–MIR ellipsometric data in SpectraRay/4, using a modified Sernelius formula. $m^*_c(m_e)$ obtained using Eq. (\ref{eq26}). Optical bandgap (\( E_g^{\text{opt}} \)), Urbach energy ($E_U$), and exciton binding energy ($E_b$) retrieved from modeling the imaginary part of the dielectric function (\( \varepsilon_2 \)) using the Elliott-Band Fluctuations optical dispersion model. The first two rows correspond to the nondegenerate limit, thus are excluded from the band description. Nevertheless, the latter are retained to assess the intrinsic bandgap ($E_{g0}$) obtained from the global fit.}

\begin{ruledtabular}
\begin{tabular}{cccccccc}
 $n_{\text{e}}$ (cm$^{-3}$) & $\mu_{\text{e}}$ (cm$^2$/V$\cdot$s) & $\omega_p$ (eV) & $\omega_{\tau}$(eV) & $m^*_c(m_e)$ & $E_g^{\text{opt}}$ (eV) & $E_U$ (meV) & $E_b$ (meV) \\
\hline
$6.61 \times 10^{16}$ (16)  & 0.96 (23)  & -   & - & -  & 3.45 (1)  & 47.03 (403)  & 49.12 (299) \\
$1.56 \times 10^{17}$ (02)  & 3.87 (06)   & -   & -  & -  & 3.50 (1)  & 63.61 (348)  & 58.92 (193) \\
$2.55 \times 10^{19}$ (22)  & 3.32 (92)   & 0.609(3)  & 0.285(2)  & 0.095(9) & 3.50 (1)  & 105.78 (214)  & 39.73 (301) \\
$4.14 \times 10^{19}$ (02)  & 6.67 (32)  & 0.729 (2)  & 0.286 (2) & 0.107(5) & 3.55 (1)  & 70.50 (1)     & 39.14 (211) \\
$5.48 \times 10^{19}$ (01)  & 4.00 (82)  & 0.763 (5)  & 0.306 (26) & 0.130(4) & 3.54 (1)  & 105.77 (1)    & 45.28 (136) \\
$8.03 \times 10^{19}$ (07)  & 5.44 (05)   & 0.865 (1)  & 0.145 (1)  & 0.148(22) & 3.64 (1) & 110.81 (288) & 57.69 (319) \\
$1.08 \times 10^{20}$ (01)  & 19.12 (24) & 0.960 (2)  & 0.235 (2)  & 0.122(1) & 3.69 (1)  & 98.95 (174)  & 46.96 (144) \\
$1.16 \times 10^{20}$ (03)  & 6.76 (15)  & 0.987 (1)  & 0.288 (2)  & 0.164(5) & 3.77 (1)  & 113.39 (284) & 48.37 (357) \\
$1.18 \times 10^{20}$ (01)  & 6.97 (76)  & 0.987 (1)  & 0.128 (1)  & 0.169(25) & 3.68 (1) & 91.32 (1824) & 55.39 (461) \\
$1.30 \times 10^{20}$ (02)  & 4.76 (05)   & 0.964 (4)  & 0.302 (1)  & 0.193(1) & 3.82 (1) & 116.51 (177) & 51.65 (242) \\
$1.35 \times 10^{20}$ (02)  & 10.60 (20) & 1.064 (1)  & 0.132 (1)  & 0.164(4) & 3.77 (1) & 113.58 (1628)& 52.71 (073) \\
$1.40 \times 10^{20}$ (02)  & 10.45 (16) & 0.978 (1)  & 0.224 (1)  & 0.196(4) & 3.84 (1)  & 116.92 (214) & 49.89 (206) \\
$1.41 \times 10^{20}$ (07)  & 10.64 (54) & 1.136 (1)  & 0.229 (1)  & 0.191(5) & 3.85 (1)  & 89.01 (167)  & 54.78 (375) \\
$1.45 \times 10^{20}$ (04)  & 11.15 (32) & 1.068 (1)  & 0.162 (5)  & 0.180(3) & 3.87 (2)  & 101.47 (647)& 41.99 (314) \\
$1.51 \times 10^{20}$ (02)  & 9.96  (16)  & 1.164 (1)  & 0.289 (3)  & 0.154(2) & 3.80 (1)  &  148.24 (319) & 50.04 (238) \\
$1.53 \times 10^{20}$ (06)  & 10.76 (45)  & 1.030 (1) & 0.172 (3)  & 0.196(8) & 3.85 (1) & 115.53 (707) & 48.37 (266) \\
$1.59 \times 10^{20}$ (03)  & 12.25 (24)  & 1.054 (1) & 0.248 (1)  & 0.197(2) & 3.83 (2) & 130.94 (1109) & 47.37 (366) \\
$1.70 \times 10^{20}$ (02)  & 11.38 (11) & 1.066 (1)  & 0.270 (3)   & 0.198(2) & 3.91 (3) & 82.15 (140)  & 39.97 (292) \\
$1.71 \times 10^{20}$ (06)  & 10.62 (35)  & 1.088 (3) & 0.206 (7)  & 0.199(2) & 3.91 (1) & 129.80 (322)  & 52.41 (396) \\
$1.86 \times 10^{20}$ (02)  & 12.08 (38)  & 1.100 (3) & 0.207 (2)  & 0.202(9) & 3.88 (1) & 146.27 (134)  & 57.11 (343) \\
$2.130 \times 10^{20}$(005) & 12.89 (02) & 1.224 (1) & 0.279 (3)  & 0.203(1)& 3.83(1) & 99.63 (132)   & 35.43 (129) \\
\end{tabular}
\end{ruledtabular}
\end{table*}

\section{EXPERIMENTAL DETAILS}
AZO thin films were synthesized by RF dual-magnetron sputtering in a confocal target configuration using high-purity ZnO ($99.99\%$
) and Al ($99.9\%$
) targets as host oxide and dopant source, respectively. Depositions were carried out on one-side polished Si ($100$)  and two-side polished fused silica substrates. During deposition, the substrate holder was continuously rotated to promote uniform film thickness and elemental composition homogeneity, while being  actively cooled using a closed cycle water system maintained at $\sim 12^\circ\mathrm{C}$ to minimize sputtering-induced substrate heating. Depositions were performed in an Ar atmosphere with a controlled gas flow of \qty{30}{\sccm}, under a working pressure of \SI{1e-2}{\milli\bar} after evacuation of the chamber to a base pressure below \SI{2.0e-6}{\milli\bar}. To systematically tune the carrier concentration, two complementary processes were employed. First, the RF power applied to the Al target was varied between \qtylist[list-units = single]{12;20}{\watt}, while the ZnO target power was kept constant at \qty{80}{\watt}. Second, post-deposition thermal annealing treatments  were performed in air and/or ultra-high-purity Ar atmospheres (\SI{\sim4.0e-1}{\milli\bar}) at temperatures ranging from \qtyrange[range-units = single]{200}{600}{\degreeCelsius} to further modulate the carrier concentrations.
These combined procedures allowed systematic tuning of the carrier concentration over the range from \qtyrange[range-units = single]{6.6e16}{2.13e20}{\cm^{-3}}, spanning both the nondegenerate and degenerate regimes across the Mott critical concentration. Additional details regarding the sputtering system and process control are provided elsewhere \cite{Enrique_2025,Llontop2023}.

X-ray diffraction (XRD) patterns were obtained using a Bruker D8 Focus diffractometer in Bragg–Brentano configuration. All samples exhibited polycrystallinity with a  preferential orientation along the (002) plane (Fig. S1(a) SM), characteristic of the hexagonal wurtzite structure of ZnO [PDF 00-036-1451]. Complementary scanning electron microscopy (SEM) performed with a ZEISS Merlin instrument equipped with a field-emission source revealed well-defined grains with columnar growth perpendicular to the  substrate surface, while smaller crystallites near the Si substrate displayed random orientations (Fig. S1(b)SM). These microstructural features, including finite mosaic spread, grain boundary misorientation, and slight tilts among adjacent crystallites, tend to average out directional differences. As a result, the films can be regarded as exhibiting minor anisotropy, which provides a reasonable basis for applying isotropic relations in band structure evaluation and accounts for the good agreement observed in the fitting of the optical data. Energy-dispersive X-ray spectroscopy (EDS)  performed using a FEI Quanta 650 SEM showed that the aluminum content in the AZO films ranged from \qtyrange[]{1.03}{4.63}{\atpercent}.

Variable angle broadband spectroscopic ellipsometry (VASE)  measurements were conducted using a SEN Research 4.0 Ellipsometer in the spectral range \qty{200} to \qty{3500}{\nm}, with incident angles from \ang{50} to \ang{70} in increments of \ang{5}. Complementary optical transmittance (T) spectra were acquired at normal incidence over the same wavelength range using the same system. Reflectance (R) data were obtained using a Perkin Elmer Lambda 950 double-beam UV–VIS spectrophotometer equipped with an integrating sphere of diameter \qty{150}{\mm} , covering the \qty{200}{\nm} to \qty{2500}{\nm} range. A three-layer (roughness, film, substrate) optical model was used for the optical data fitting, yielding film thicknesses between \(220 \pm 0.3\,\mathrm{nm}\) and \(375 \pm 0.3\,\mathrm{nm}\). Surface roughness was modeled using the Bruggeman effective medium approximation (EMA) with a fraction of \( 50\% \) air inclusion, resulting in roughness values ranging from \(8 \pm 0.6\,\mathrm{nm}\) to \(22 \pm 0.5\,\mathrm{nm}\). A dispersionless self-consistent point-by-point method was used to retrieve the complex refractive index  $\widetilde{n}$ and the corresponding complex dielectric function, \( \widetilde{\varepsilon} = \varepsilon_1 + i\varepsilon_2\) \cite{Enrique_2025,Guerra17}. Furthermore, Kramers–Kronig (KK) consistency was verified for all retrieved $\widetilde{n}$ curves by leveraging several redundant optical measurements to constrain $\widetilde{n}(\lambda_i)$ at each measured photon wavelength $\lambda_i$. The resulting dielectric functions were subsequently modeled to extract relevant optical parameters  from the free carrier and fundamental absorption regions, as detailed in Ref.~\cite{Enrique_2025}. Representative examples of the point-by-point spectral fitting for samples with low, intermediate, and high carrier concentrations, are shown in SM Fig.~S2.

Carrier concentration ($n_e$) was obtained from  Hall effect measurements performed in the Van der Pauw configuration at room temperature. The setup employed a Keithley 6485 picoammeter and 2182A nanovoltmeter integrated with a 3765 Hall effect card used for the switching configuration. $40\text{ mm}$  tungsten-rhenium alloy probe tips were used as mechanical electrical contacts. The ohmic behavior was tested before each measurement. An Instec M06T Hall effect stage equipped with a HCP621G-PMH Instec Hall probe with a reversible 0.42 T magnetic field was used. 
For the samples with lower carrier concentrations (nondegenerate samples), resistivity and Hall effect measurements were additionally performed using a Lakeshore Model 8404 AC/DC Hall Effect measurement system. This system was controlled through a LabVIEW-based routine enabling variation of both the injected current ($I$) and magnetic field intensity ($B$). The Hall coefficient ($R_H$) was determined from the slope of the linear dependence of the Hall voltage ($V_H$) as a function of $I \cdot B$. Subsequently,  Hall mobility ($\mu_e$) was calculated using the measured carrier concentration ($n_e$) and resistivity values, while uncertainties were estimated through error propagation  arising from the fits. Across the investigated carrier concentration regime, the Hall-derived concentrations are taken to approximate the effective free-electron population in the conduction band, consistent with the degenerate nature of the samples.

\section{Results and discussion}

\begin{figure*}[t!]
 \centering
\includegraphics[width=\textwidth]{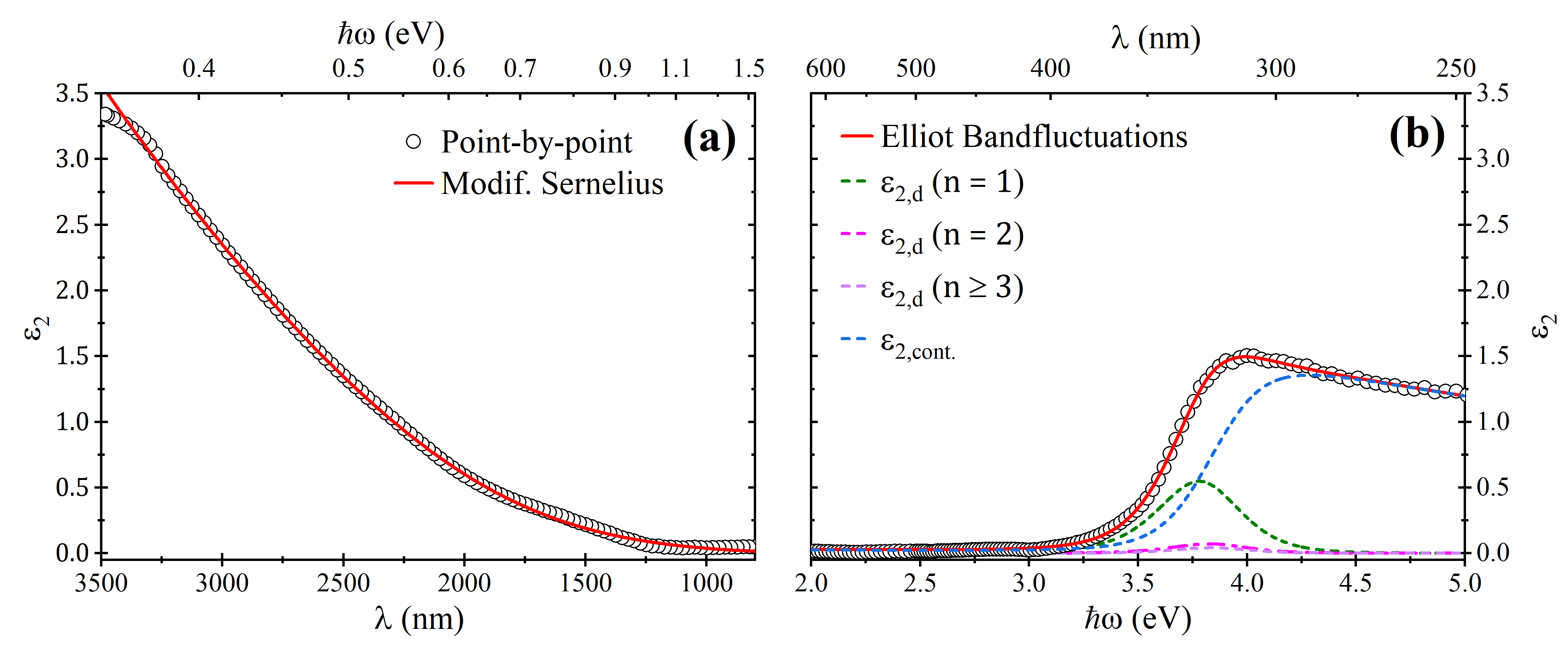}
\caption{Imaginary part of the dielectric function \((\varepsilon_2)\) extracted using a model-independent point-by-point method through the simultaneous fitting of spectroscopic ellipsometry, transmittance, and reflectance data for an AZO thin film with a carrier concentration of \qty{1.53e20}{\cm^{-3}}. Fits (red solid lines) of the free carrier absorption (a) and fundamental absorption (b) spectral regions, using the modified Sernellius formula and the Elliott–Band Fluctuations optical dispersion model, are shown, respectively. Additionally, the excitonic continuum and discrete absorption contributions are also depicted in (b) as dashed lines. Here, the discrete excitonic peaks are denoted as $\varepsilon_{2,d}\,(n=1)$, $\varepsilon_{2,d}\,(n=2)$, and $\varepsilon_{2,d}\,(n\geq3)$ for the first, second, and higher-order excitonic states, respectively, while the continuum contribution is abbreviated as $\varepsilon_{2,\mathrm{cont.}}$. Point-by-point data density is reduced for viewing purposes only.}
\label{Figure2}
\end{figure*}

The free carrier absorption was modeled  through the complex dielectric function $\widetilde{\varepsilon}$, using a modified Sernelius optical dispersion formula. From the latter, the plasma ($\omega_p$) and damping ($\omega_\tau$) frequencies  were retrieved,  following the procedure  detailed in Ref.~\cite{Enrique_2025}. Further,  the Ioffe–Regel criterion ($k_{\text{F}} l > 1$) was also evaluated using the optical relaxation time obtained from $\omega_\tau$ \cite{Hussey2004,Laad2024}. While Hall-based values, which are strongly affected by macroscopic transport and microstructural scattering, the optically obtained $k_F\ell $, where $\ell $ the optical mean free path, remain well above unity ($k_F l \approx 2.3\text{--}8.8$) over the entire carrier concentration range analyzed, supporting the applicability of the degenerate carrier band description to the optical response \cite{Fujiwara,KEllmer}. The fundamental absorption was modeled using a modified Elliott formula that incorporates overlapping contributions from excitonic and tail states \cite{kevin25,Guerra19}. This approach accurately reproduces the observed features in the experimental $\varepsilon_2$ curves,  enabling the reliable extraction of the optical bandgap ($E_g^{\mathrm{opt}}$), Urbach energy ($E_U$), and exciton binding energy ($E_b$).  Additionally, the variation of excitonic broadening with carrier concentration is consistent with the screening of the electron–hole Coulomb interaction
in a degenerate semiconductor \cite{Makino_2005,Makino2002,Enrique_2025, Schleife2011}, as illustrated in Sec.~S4. As demonstrated in our previous work, neglecting free excitonic contributions and tail-states absorption, can lead to a systematic underestimation of $E_g^{\mathrm{opt}}$ by approximately \qty{0.25}{eV} \cite{Enrique_2025,Asghar}. Such deviations in the optical bandgap propagate directly into the evaluation of the BMS, BGR, and consequently the total bandgap energy shift, thereby biasing subsequent optoelectronic analyses. Representative experimental \( \varepsilon_2 \) spectra along with their corresponding fits across both the free carrier and fundamental absorption regions are presented in Fig.~\ref{Figure2}. The evolution of these spectral components with increasing carrier concentration is illustrated in Fig.~S2. The key optical and electronic transport parameters of the investigated AZO thin films obtained using this approach are summarized in Table~\ref{table1}.

\begin{figure*}[t]
    \centering
    \includegraphics[width=\textwidth]{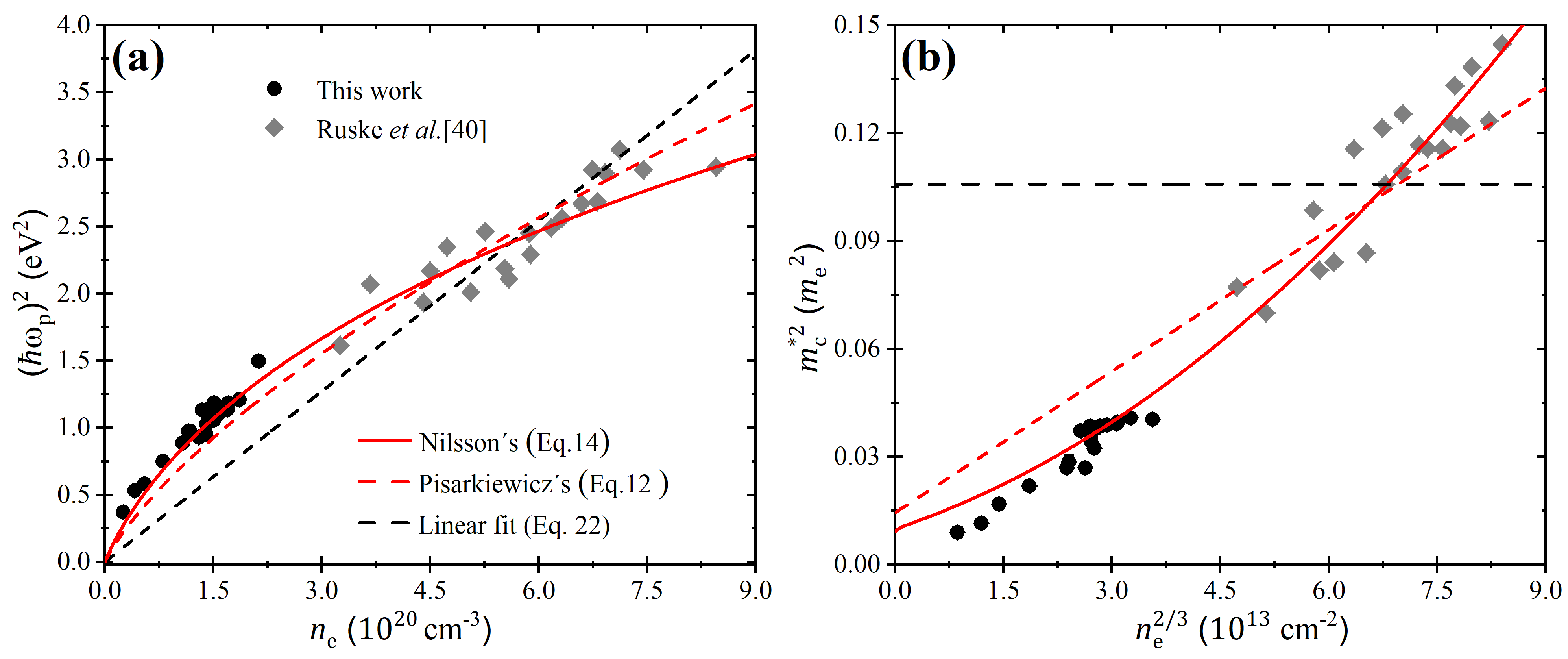}
    \caption{ Squared plasma energy $(\hbar \omega_p)^2$ extracted from the free carrier absorption fit, plotted  as a function of  the carrier concentration, $n_e$ (a). Squared electron effective mass ${m_c^*}^2$ calculated using Eq.~(\ref{eq26})  as a function of  $n_e^{2/3}$ (b). In both graphs, black dashed lines represent the best linear fits obtained from Eq.~(\ref{eq26}) assuming a constant effective mass, whereas the red dashed and solid lines were generated using the  nonparabolic parameters $m_{c_0}^\ast$ and $C$ obtained from the global fit described in Sec.~II Eq.~(\ref{eq23}), combining our experimental data listed in Table~\ref{table1} together with the dataset of Ruske \textit{et al.}, Table~S2. Band nonparabolicity is modeled using the Nilsson (solid) and Pisarkiewicz (dashed) relations, Eqs.~(\ref{eq15}) and~(\ref{eq13}), respectively.
    }
    \label{Figure3}
\end{figure*}

\begin{figure*}[t]
    \centering
    \includegraphics[width=\textwidth]{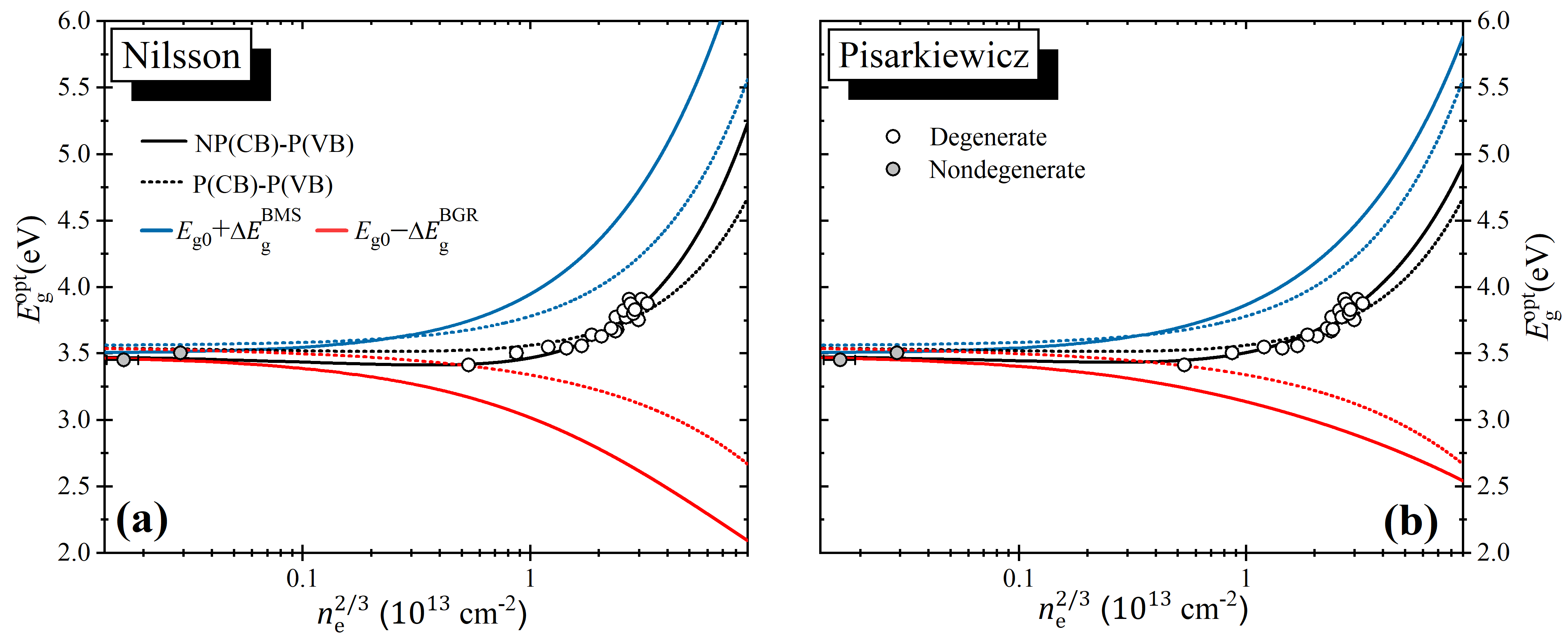}      
    \caption{ Best fitted curves (black) of the optical bandgap, \(E_g^{\mathrm{opt}}\)  as a function of  \(n_e^{2/3}\)  following Eq.~(\ref{eq3}). Individual contributions from the Burstein-Moss shift (blue curves) and bandgap renormalization (red curves), calculated within the RPA, are also shown. Results considering a nonparabolic conduction band with parabolic valence band (NP(CB)-P(VB), solid lines) and both parabolic conduction and valence bands (P(CB)-P(VB), dotted lines) are presented for comparison purposes. Nonparabolicity is described with the  Nilsson (a) and Pisarkiewicz (b) relations Eqs.~(\ref{eq15}) and ~(\ref{eq13}) respectively. Best fitted curves are obtained after the global fit described in the Methodology section Eq.~(\ref{eq23}), combining the experimental values listed in Table~\ref{table1} together with the dataset of Ruske \textit{et al.}, Table~S2. 
}
 \label{Figure4}     
\end{figure*}

 \begin{figure*}[t]
    \centering
    \includegraphics[width=\textwidth]{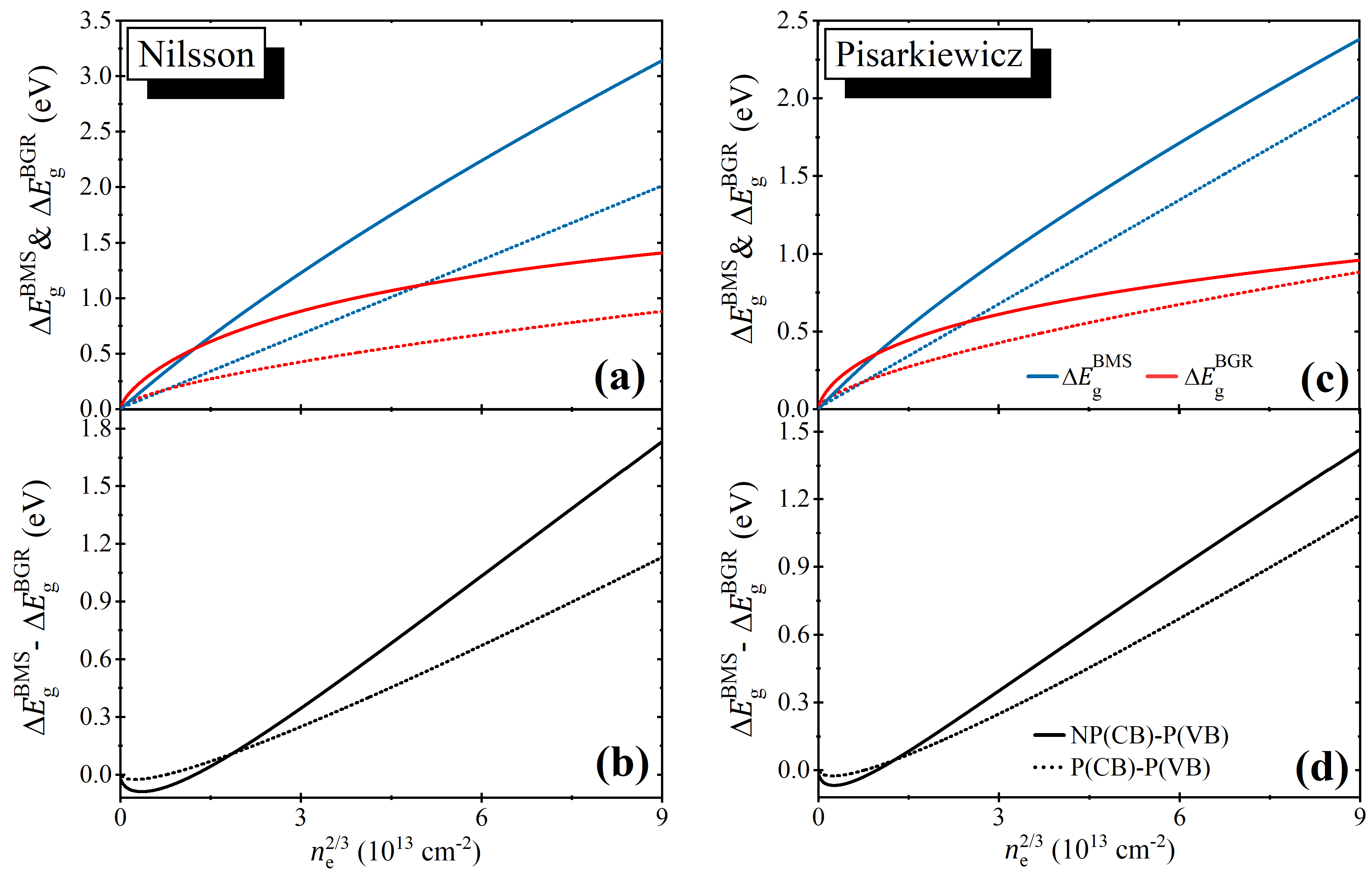}      
    \caption{Burstein–Moss shift (blue), and bandgap renormalization (red)  as a function of  \(n_e^{2/3}\) after Nilsson (a) and Pisarkiewicz (c), along with the total bandgap energy shift ($\Delta E_g^{\text{BMS}} - \Delta E_g^{\text{BGR}}$) after Nilsson (b) and Pisarkiewicz (d). Shifts within parabolic conduction and valence bands (P(CB)–P(VB)) are shown as dotted lines for comparison purposes. All calculations have been performed considering RPA self-energy terms. Equivalent curves considering PPA self-energy terms can be found in, Fig. S7. 
}
 \label{Figure5}     
\end{figure*}
Plasma frequency dependency with the carrier concentration can be assessed through the squared plasma energy relation, 
\begin{equation}
(\hbar\omega_p)^2 = \frac{\hbar^2 e^2 n_e}{\varepsilon_0 m_c^*}.
\label{eq26}
\end{equation}
The conduction effective mass ${m_c^*}$ can be directly calculated by isolating it from  Eq.~(\ref{eq26}) and substituting the optically determined plasma frequency together with the carrier concentrations measured via the Hall effect. Fig.~\ref{Figure3} shows the  square of plasma energy dependence on carrier concentration, plotted as $(\hbar\omega_p)^2$ versus $n_e$, as well as ${m_c^*}^2$ versus $n_e^{2/3}$. Plasma energy data reported by Ruske \textit{et al.}~\cite{RUSKE2009} is included showing close agreement with the present results and enabling the extension of the analysis to higher carrier concentration ranges. The data set  from Ruske \textit{et al.} is available in Table S2 of the SM.

Within the parabolic band approximation $(\hbar\omega_p)^2$ plotted against $n_e$ follows a linear relation passing through the origin, with a slope inversely proportional to a constant ${m_c^*}$. Whereas, for the nonparabolic case, ${m_c^*}$ continuously increases with $n_e$, causing $(\hbar\omega_p)^2$ to deviate from linearity.
For carrier concentrations up to \qty{2.13e20}{cm^{-3}}, a linear fit yields $m_{c_0}^\ast = 0.18\text{ }m_e$, consistent with the reported values for ZnO \cite{Jianguo,Mott1961Metallic}. Nevertheless, over this same carrier concentration range, ${m_c^*}^2$  is not constant, and the parabolic model fails to reproduce the experimental data, as shown in Fig.~\ref{Figure3}(b). Furthermore, extending the parabolic fit across the entire carrier concentration range, as shown in Fig.~\ref{Figure3}, results in an apparent effective mass of $0.32\,m_e$, highlighting the limitations of the parabolic band approximation in heavily doped regimes. As depicted in Fig.~\ref{Figure3}, $(\hbar\omega_p)^2$ deviates systematically from linearity above the Mott carrier concentration, $n_c \approx 4.85 \times 10^{19}\,\mathrm{cm^{-3}}$ \cite{Roth,Jianguo}, signaling the onset of pronounced conduction band nonparabolicity. The approximately linear dependence of ${m_c^*}^2$ on $n_e^{2/3}$, suggests that the previously presented nonparabolic dispersion models of Pisarkiewicz Eq.~(\ref{eq13}) and Nilsson Eq.~(\ref{eq15}), are applicable. Although the Pisarkiewicz relation reproduces the general trend, an independent fit of this model, to $(\hbar\omega_p)^2$ as a function of $n_e$ yields unphysical nonparabolic parameters, $m_{c0}^{}=0.02\,m_e$ and $C^{-1}=0.2\, \mathrm{eV}$, which deviate significantly from the expected value of $\sim \nicefrac{1}{E_{g0}}$. In contrast, an independent fit of the same data set using  Nilsson’s relation directly, yields physically meaningful parameters $m_{c0}^{*}=0.08\,m_e$ and $C^{-1}=3.57\,\mathrm{eV}$, and provides a markedly improved description of the experimentally observed behavior at room temperature. A detailed comparison of the independent direct fits can be found in Sec. S5 of the SM. The improved agreement may arise from the additional physical effects incorporated in Nilsson’s model, including finite temperature  occupation  effects through Fermi–Dirac statistics together with disorder- and impurity-induced band tailing. As a result, this model captures both intrinsic band filling and extrinsic broadening mechanisms, leading to a  physically consistent representation of the band nonparabolicity. 

The wide variation of reported effective masses and nonparabolicity parameters in AZO and other wide bandgap oxide thin films, including ITO, GZO, and IZO \cite{RUSKE2009,Young,Kim_2008,ABDOLAHZADEHZIABARI2012,Feneberg16,Bikowski,Fujiwara,Papadimitriou,Jianguo,Walsh,Jia,Saw}, likely reflects differences in fitting procedures, carrier-concentration ranges, and measurment techniques. Together, these studies highlight the need to
report both conduction band edge effective mass $m_{c0}^{*}$ and the nonparabolicity parameter $C$. To overcome this ambiguity, we employ the global fitting strategy described in Sec. II D, in which the band parameters are constrained simultaneously by the fundamental absorption and free carrier spectral responses. The analysis incorporates a finite hole effective mass in the combined treatment of the BMS and BGR, providing independent constraints beyond the free carrier response alone. The best fitted curves after the global fitting are shown in Figs.~\ref{Figure3} and \ref{Figure4}. Whereas the best fitted parameters are shown in Table~\ref{table2}.  The extracted $m_{c0}^{*}$ is significantly lower than the previously reported values, while the corresponding $C$ value lies within the expected dependence $\sim \nicefrac{1}{E_{g0}}$. Importantly, the proper combination of $m_{c0}^{*}$ and $C$, obtained through Nilsson’s relation, provides a physically meaningful description of the experimental data. Within the present analysis, the reduced $m_{c0}^{*}$ value may reflect modifications of the band edge electronic structure driven by impurity-induced tail states and bandgap narrowing, alongside intrinsic conduction band nonparabolicity \cite{Sernelius,Jia}. This interpretation is consistent with our bandgap renormalization analysis depicted in Figs.~\ref{Figure4} and \ref{Figure5}. Further, the optical response considered here does not isolate a specific crystallographic component of the effective mass tensor, and therefore represents a mixed response of the parallel and perpendicular components, which can also result in a smaller effective masses \cite{Kim_2008}. Ultimately, the magnitude of $m_{c0}^{*}$, depends on both the carrier concentration range probed and the extraction technique applied. Consequently, this value should not be interpreted as a universal, isotropic electron effective mass for ZnO. These optimized parameters serve as the foundation for the subsequent analysis of the BMS, BGR, and hole effective mass.
 
Fig.~\ref{Figure4} compares the competing BMS and BGR shifts  calculated using two distinct band dispersion approximations i.e., parabolic conduction and valence bands (P(CB)–P(VB)) and our  proposed nonparabolic conduction with parabolic valence bands (NP(CB)–P(VB)), following the methodology  detailed in Sec.~II. Contrasting NP(CB)–P(VB) approximation with widely used simplified approaches reveals pronounced deviations  in the predicted shifts, with differences around
$0.3$–$0.6~\mathrm{eV}$. Specifically, adopting both parabolic conduction and valence bands (P(CB)–P(VB)) as from Fig.~\ref{Figure4} or nonparabolic conduction band without including a finite valence band, as detailed separately in Sec.~S8 of SM, leads to a systematic underestimation of  approximately $0.4$–$0.6~\mathrm{eV}$ in the BMS and up to $\sim 0.5~\mathrm{eV}$ in the BGR producing a net bandgap shift error of approximately $0.3$–$0.6~\mathrm{eV}$. Notably, these deviations directly explains the widespread discrepancies in literature-reported values, which frequently stem from omitting valence band self-energy corrections~\cite{Feneberg16,Walsh,Feneberg14,ABDOLAHZADEHZIABARI2012}, calculating the BMS relative to the conduction band edge alone~\cite{Feneberg16,Feneberg14}, or relying on uncorrected parabolic models in carrier-dependent calculations ~\cite{Jia,Whalley,Walsh,Jianguo}.  Beyond these numerical deviations, simplified models conceal key physical mechanisms. Although their predicted net shifts can be forced to superficially match experimental trends, they fail to capture the dominance of BGR below $n_c$, rendering them unable to reproduce the underlying many-body contributions. In contrast, the NP(CB)–P(VB) approximation within the RPA incorporating Nilsson's nonparabolic effective mass  relation, resolves this ambiguity by explicitly separating BMS and BGR contributions, closely matching the experimental carrier concentration dependence, shown in Fig.~\ref{Figure5}. 

Building upon this separation, the present analysis provides an alternative experimentally constrained estimate of the characteristic Mott concentration \(\sim n_c\) derived directly from the BMS–BGR crossover, rather than relying solely on the conventional $n_e^{1/3}a^{*}\approx K$ criterion, which is based on simplified screening, and relies on an empirical proportionality constant $K\sim0.18$-$0.37$ \cite{Jianguo,GUPTA198933}. Within the RPA-Nilsson combination, this BMS-BGR crossover occurs at $n_e^{2/3}$ $\sim1.33\times10^{13}$ cm$^{-2}$ (i.e., $n_e = 4.85\times10^{19}\,\mathrm{cm}^{-3}$),  and at a comparable concentration for the PPA-Pisarkiewicz combination, values that are consistent with reported Mott concentrations for ZnO-based systems \cite{Jianguo,Roth}. Furthermore, its agreement with the concentration threshold where $(\hbar\omega_p)^2$ vs. $n_e$ deviates from the parabolic band approximation  provides an independent validation of our modeling approach. Around this characteristic concentration, both BMS and BGR increase with carrier concentration, BGR dominates below the crossover, led primarily by valence band electron–impurity interactions, whereas BMS becomes increasingly dominant at higher concentrations. This crossover thereby marks the regime where band filling and many-body renormalization balance, while its extension toward lower concentrations defines the lower bound for the validity of the degenerate-carrier approximation.

In addition, the choice between RPA or PPA  within the presented methodology is also relevant. For instance, the P(CB)–P(VB) approximation within the RPA underestimates the total bandgap energy shift by approximately \qty{0.35}{eV} towards the higher carrier concentration, while a systematic offset of approximately \qty{0.15}{eV} is observed between the RPA  and PPA  self energy terms for both the P(CB)–P(VB) and NP(CB)–P(VB) approximations, indicating a general tendency of the PPA formalism to overestimate the total energy shift as shown in SM. In contrast, under the Pisarkiewicz’s nonparabolic relation, PPA underestimates the renormalization shift, relative to RPA. This contrasting behavior reflects the fundamentally different ways in which the single-pole screening approximation couples to the underlying electronic band structures. Within the Nilsson relation, the PPA enhances conduction band contributions, whereas within the Pisarkiewicz’s relation it does not adequately reproduce the redistribution of spectral weight among the available electronic states. By comparison, the RPA formalism, through the integration over the full continuum of electron–hole excitations, incorporating full frequency-dependent screening and interband coupling, provides a self-consistent description of the BGR self energy terms. Collectively, these results demonstrate that both weak- and strong-localization effects contribute significantly to the screened dielectric function across the investigated carrier concentration range. The overall consistency achieved within the RPA  further supports the reliable extraction of carrier effective masses, summarized in Table~\ref{table2}, while properly accounting for conduction band nonparabolicity. Additional comparisons testing the present approach against commonly adopted models, including PPA-based fitting, BMS and BGR calculations neglecting valence band contributions, RPA and PPA treatments of BGR, along with the Nilsson and Pisarkiewicz nonparabolic band relations, are provided in SM Secs. S7 and S8 with Figs. S6, S7 and S8.

\begin{table}[t]
    \centering
    \caption{Best fitted parameters obtained under the global fit using the nonparabolic conduction band with parabolic valence band  (NP(CB)–P(VB)) and parabolic conduction and valence bands (P(CB)–P(VB)) approximations within the RPA.}
    \begin{ruledtabular}
    \begin{tabular}{lccc}
    Parameter & Pisarkiewicz & Nilsson & Parabolic bands \\
    \hline
    $m_{c_0}^*(m_e)$ & 0.12(1) & 0.10(1) & 0.32(1) \\
    $C~(\mathrm{eV}^{-1})$ & 0.75(2) & 0.26(1) & -- \\
    $E_{g_0}~(\mathrm{eV})$ & 3.50(3) & 3.50(4) & 3.55(2) \\
    $m_{v_0}^*(m_e)$ & 0.32(4) & 0.32(2) & 0.33(2) \\
    \end{tabular}
    \end{ruledtabular}
    \label{table2}
\end{table}

The hole effective mass in AZO/ZnO remains a critical yet underexplored parameter for carrier dynamics, particularly in the degenerate-carrier regime. In many prior studies, it is either neglected or assumed to be infinitely large \cite{Feneberg16,Asghar}, while others rely on simplified models with a constant electron effective mass \cite{GUPTA198933,Sernelius}, failing to capture the complexity introduced by carrier-dependent nonparabolicity. 
Experimental determinations of the hole effective mass in doped-ZnO are limited, with only a few studies based on cyclotron resonance and photoluminescence measurements \cite{ZnOValenceBand1999,Shi,Asghar,OSHIKIRI2001}. A systematic understanding of the influence of impurity incorporation on the hole effective mass in doped ZnO is therefore still lacking. Similarly, theoretical approaches such as  $k\cdot {p}$ formalism, evaluate the mean hole effective mass  by incorporating strain‑dependent energy levels near the $\Gamma$ point. Considering the valence band dispersion along the dipole‑allowed transition direction, the effective masses of ~$0.48\,m_e$ for both HH (Heavy hole) and LH (Lighthole) states and ~$0.26\,m_e$ for CH (Chiral hole) states have been reported for ZnO, corresponding to an average value of ~$0.40\,m_e$ \cite{Tripathi2020,ShadangiIOP2016,Klingshirn,klingshirn2010}. For ZnO single crystals, crystallographic directions dependent hole effective masses of ~ $0.66\,m_e$, $0.64\,m_e$, $0.53\,m_e$ have been reported, corresponding to an average of ~$0.61\,m_e$ \cite{SYRBU2004}. Theoretical estimates based on first-principles calculations provide additional insight into the valence band dispersion, however, the available reports are limited to ZnO \cite{Sikam_2020,OSHIKIRI2001,Lambrecht}.  The present analysis yields an average hole effective mass of $0.32\,m_e$, close, yet distinctly lower than, these reported values. This deviation can be attributed primarily to carrier-induced many-body effects associated with Al doping, where the increased free electron concentration modifies the valence band curvature through dielctric screening and the resulting bandgap renormalization. These effects can modify the valence band curvature and therefore the  hole effective mass.

This distinction is particularly important when compared with previous results based on the total bandgap-shift relation $\Delta E_g=\Delta E_g^{\mathrm{BMS}}-\Delta E_g^{\mathrm{BGR}}$ \cite{jain_1990}. For instance, Esmaliei \cite{Asghar}, reported anomalously large negative hole effective masses, ranging from approximately $-3.74\,m_e$ \ \text{to} \ $-1.07\, m_e$, when applying a similar analysis to ZnO and degenerate AZO. Such values lie well outside the physically expected range and indicate the sensitivity of the extracted hole effective mass to the assumptions entering the BMS and BGR terms. A negative hole effective mass of this magnitude would imply an effective valence- and conduction-band curvature inconsistent with the conventional band edge picture, thereby introducing substantial uncertainty into the resulting BGR correction. The limitations of this approach and their consequences for the extracted effective mass have been discussed in detail by Gupta \textit{et al.} \cite{GUPTA198933}. 

Our findings establish a consistent methodology, where the  hole effective mass in AZO is finite and experimentally accessible. Additionally, by demonstrating that Al incorporation modifies the hole effective mass, we highlight the need for systematic studies across varying impurity concentrations and types.

\section{Summary and Conclusions}

We have developed an unified, self-consistent framework to model the optical bandgap evolution in Al-doped ZnO across a wide carrier concentration range by incorporating band nonparabolicity, the Burstein–Moss shift (BMS), and bandgap renormalization (BGR). This methodology accurately reproduces experimental trends for both the optical bandgap shift and the electron effective mass, the latter exhibiting a steep increase from $0.095\,m_e$ to $0.42\, m_e$  over the  carrier concentration range of \qtyrange[range-units = single]{2.55e19}{8.45e20}{\cm^{-3}}, highlighting clear deviations from parabolic behavior.

A global optimization simultaneously fitting the carrier-concentration dependence of the optical bandgap and squared plasma energy yields a band edge electron effective mass ($ m_{c_0}^\ast$) of $0.10\,m_e$, a nonparabolicity parameter (\( C \sim 1/E_{g0} \)) of  $0.26 \,\text{eV}^{-1}$ and an intrinsic bandgap of \qty{3.5}{eV}, aligning closely with the established value for ZnO (\qty{3.4}{eV}). The close correspondence between $C$ and $1/E_{g0}$ further corroborates the physical self-consistency of the extracted parameters. The optical bandgap, obtained from fitting the imaginary dielectric response \(\varepsilon_2\) with excitonic and disorder-induced tail state contributions, exhibits a pronounced blue shift of $\sim 0.45\,\mathrm{eV}$. This shift is quantitatively captured by combining BMS and BGR effects within Kane’s dispersion relation and using a frequency-dependent dielectric response for many-body screening.

Comparison of many-body approximations shows that while the plasmon pole approximation (PPA) overestimates bandgap shifts in the presence of nonparabolicity, the random phase approximation (RPA), incorporating dynamic screening and interband coupling, provides a more accurate description of the BGR, differing from PPA by up to $\sim 0.15\,\mathrm{eV}$. The analysis further reveals that BGR dominates below the Mott carrier concentration ($n_c$), while BMS becomes increasingly significant above it. Further, the framework yields a mean hole effective mass of $0.32\, m_e$, a parameter that, to the best of our knowledge remains poorly constrained in ZnO-based systems. 

Ultimately, these findings confirm that reliably modeling carrier-induced bandgap modulation requires the simultaneous integration of nonparabolicity, BMS, BGR, and finite valence-band dispersion. We believe the proposed unified methodology establishes a versatile tool for characterizing transparent conducting oxides and degenerate transparent semiconductors, offering a solid foundation for evaluating mobility-limiting scattering mechanisms to optimize optoelectronic device performance.

\section*{Acknowledgements}
This research was funded by the Air Force Office of Scientific Research (AFOSR), Grant No. FA9550-25-1-0006. S. Mishra acknowledges the support from the Pontificia Universidad Católica del Perú (PUCP) for the PhD scholarship grant No. DOC-2023-003. F.Bravo acknowledges the support from Vice Chancellorship of research CAP Grant No. PI-0888. L. A. Enrique acknowledges support from PUCP through the PhD scholarship grant No. DOC-2025-007. The authors acknowledge the Center for Materials Characterization (CAM) at PUCP and the Center for Technological, Biomedics and Environmental Research (CITBM) at UNMSM for supporting the development of this work. The authors also gratefully acknowledge  Dr. Klaus Habicht for providing access to the laboratory facilities for resistivity and Hall effect measurements at the Department of Dynamics and Transport in Quantum Materials at Helmholtz-Zentrum Berlin.

\bibliography{references}

\end{document}